\documentclass{article}
\usepackage{authblk}
\usepackage{amsmath}
\usepackage{graphicx}

\def\be{\begin{equation}}
\def\ee{\end{equation}}

\def\Xint#1{\mathchoice
   {\XXint\displaystyle\textstyle{#1}}%
   {\XXint\textstyle\scriptstyle{#1}}%
   {\XXint\scriptstyle\scriptscriptstyle{#1}}%
   {\XXint\scriptscriptstyle\scriptscriptstyle{#1}}%
   \!\int}
\def\XXint#1#2#3{{\setbox0=\hbox{$#1{#2#3}{\int}$}
     \vcenter{\hbox{$#2#3$}}\kern-.5\wd0}}

\def\dashint{\Xint-}

\title{Circular plate capacitor with different disks}

\author[1,2]{Giampiero Paffuti}
\author[1]{Enrico Cataldo}
\author[1,2]{ Alberto Di Lieto}
\author[1]{Francesco Maccarrone}
\affil[1]{Dipartimento di Fisica - Universit\'a di Pisa,  Largo Pontecorvo 7, Pisa, Italy}
\affil[2]{INFN sezione di Pisa, L.go Pontecorvo 3 Ed.  C, I-56127 Pisa, Italy}

\begin{document}
\maketitle

\begin{abstract}
In this paper we write a system of integral equations for a capacitor composed by two disks of different radii, generalizing Love's equation for equal disks. We compute the complete asymptotic form of the capacitance matrix both for large and small distances obtaining a generalization of Kirchhoff's formula for the latter case.
\end{abstract}

\section{Introduction}
The analytic calculation of the capacitance and potential coefficients of a couple of conductors of various definite shapes has a very long and distinguished history. Several eminent scientists have tackled the issue\cite{maxw1,Kirchh,thom}, and many researchers have produced specialized studies in the past decades using more and more frequently the newly available numerical resources\cite{nishi,Muro}. Due to availability of approximate solutions when the conductors are considered very far or very close from each other, much attention has been devoted to the asymptotic behavior of the parameters in the far and in the near limit as a suitable check of elaborate numerical procedures. Moreover, even an approximate  closed form is very useful in order to ascertain or conjecture on relevant and general properties of the electrostatic parameters.

It is worth to note that the long history is punctuated by definite periods of revival of the issue, often in conjunction with particular field of technical application. In the 90s of the past century considerable effort was applied to the calculation of the capacitance of planar systems in conjunction with the analysis of microstrip transmission lines\cite{nishi}.

Today, the extensive use of MEMS devices such as electrostatically driven actuators or detectors sensitive to the static charge carried by movable parts (e.g. capacitive

\maketitle

\noindent
sensors) urged an accurate knowledge of their electrical properties for selected geometries such as those occurring with parallel conducting plates or spherical-spherical and spherical-plane electrodes\cite{Muro}.
 
The potential or capacitance coefficients are the primary parameters involved in the description of the charges (voltages) variations following a deformation of the shape of conductors or a change in their separation or in the dielectric constant of the gap material. Those coefficients are also important in the investigation of the interelectrode forces arising when small charged conducting particles approach each other or one of them interacts with a close conducting surface\cite{Qin}.

In this paper we present a generalization of the  Love's integral equation\cite{Love} treating the case of disks of different diameters and giving the exact result in the near limit. To the best of our knowledge, these results are new. Moreover, we put the specific problem explored in this paper in a more general
framework: the analysis of divergences of the coefficients of capacitance when the mutual distance between conductors vanishes.

The paper is organized as follows. In section \ref{constraints} 
we put the problem in the more general framework of small distances behavior of capacitance matrix elements.
In section \ref{riassuntoLove} we give a brief review of the results for identical disks.
In section \ref{sezdiscussione} we present a summary of our results, particularly the expression of the capacity coefficients. We perform  
comparisons with large distances results and the arguments expressed in section \ref{constraints}. Section \ref{provaequazione} provides a formal proof for the basic system of integral equations which fix the solution and in section \ref{risasintotici} we perform the asymptotic small distance expansion, both for capacity and for solutions of the integral equations.

\section{Physical constraints on the coefficients $C_{ij}$\label{constraints}}

Let us consider a unique conductor, ideally composed by many approached parts, of capacitance $C_F$ which is the charge $Q_F$ acquired if the body is maintained at unitary potential. The charge is equal to the flux of electric field across the surface of the body or, more intuitively, to the number of lines of force leaving the surface and ending at infinity.
Now let us perform a small displacement, $\ell$, between two parts of the body: no force line joining the detached parts appears, if the potential of each part is held fixed, and the complex of lines acquires a small distortion. In more formal language the new total charge, $Q(\ell)$ is close to the original one, i.e. the total charge is a continuous function of $\ell$. In particular, the charge $Q_F$ is finite for $\ell\to0$, i.e. in the limit reproducing the original situation. Using the same kind of reasoning one can conclude that the charge of each part remains finite in that limit.
A more algebraic proof is obtained by observing that the sum of the individual charges $Q_i(\ell)$ gives the total charge $Q(\ell)$ and that $Q_i(\ell)$ is positive, being the sum of coefficients $C_{ij}$ for fixed $i$, see ref.\cite{maxw1} (\textsection 89). Then if $Q(\ell)$ is finite, the single $Q_i(\ell)$ cannot diverge for $\ell\to 0$.

In the particular case of a system of two conductors the above considerations amount to say
\begin{subequations}\label{limitiC}
\begin{align}
&\lim_{\ell\to 0}\left[ C_{11}(\ell) + C_{22}(\ell) + 2 C_{12}(\ell)\right] = C_F < \infty\label{limitiCa}\\
&\lim_{\ell\to 0}\, \left[C_{11}(\ell) +\, C_{12}(\ell)\right] < \infty\,;\qquad
\lim_{\ell\to 0}\, \left[C_{22}(\ell) +\, C_{12}(\ell)\right] <\infty\,.\label{limitiCb}
\end{align}
\end{subequations}
Let us stress that \eqref{limitiCb} implies a complete cancellation between possible divergences in each conductor separately, while 
\eqref{limitiCa} implies a limit value for a combination of the {\em finite} part of fringe effects.

These constraints are particularly effective in the case of parallel plates, the limit of two plates being again a plate, and even more effective for  identical parallel plates, where by symmetry
\be \lim_{\ell\to 0}\left[ C_{11}(\ell) + C_{12}(\ell)\right] = \frac12 C_F \,.\label{limitiC2}\ee
While these considerations appear rather obvious, they provide a cogent check for both analytic and numeric computation of capacitance coefficients. The kind of divergences for $\ell\to 0$ depends on the type of contact. For point-like contacts as in the case of two spheres
the divergence is logarithmic, and the relations \eqref{limitiCb} can be easily checked on the exact solutions for this case, see eg.\cite{lek}.
The situation is particularly compelling in the case of two different flat conductors of area $A$: we expect two divergent terms: a ``bulk'' divergence 
which goes like $A/\ell$, and a ``fringe'' divergence growing like $\log\ell$. On physical grounds, based on the force lines picture, one expects that the bulk divergence is ruled in general by the conductor of smaller area, while the logarithmic divergences must be essentially governed by the induction coefficient $C_{12}$. The sum of the two terms must cancel out when $C_{12}$ is combined with two {\em different} self-capacitance coefficients, $C_{11}, C_{22}$, in accordance with the constraints of eqs.\eqref{limitiC}.
We have thought that it would be instructive to verify how this would be achieved in a system disregarded in the literature, and this has been the original motivation for the study of a capacitor composed by two different coaxial disks.

\section{Parallel disks capacitor: a short review of Love's equation\label{riassuntoLove}}
The problem of a capacitor consisting of two parallel coaxial disks of radius $a$ and at distance $\ell$ has a long and rich history in the electrostatics research. The first classic result is the Kirchhoff\cite{Kirchh} formula \eqref{parametri1.5}.
A crucial step for our purposes has been done by E.R. Love\cite{Love} who, generalizing some previous results of J.W. Nicholson\cite{nick}, reduced the Laplace problem to a solution of a linear integral equation. The problem has been reformulated in a useful way using cylindrical coordinates
by I.M. Sneddon\cite{sne}. In this section we review some aspects of the problem to fix the notations and for comparison with subsequent work.
We follow the clear presentation of these results given in ref.\cite{Carlson}.

In short the solution of the general case of two equal disks at different potentials $V_1, V_2$ is reduced to a system of integral equations
\be
V_1 = F_1(t) +\int_0^1 K(t,z;\kappa) F_2(z)\,dz\,;\quad 
V_2 = F_2(t) +  \int_0^1 K(t,z;\kappa) F_1(z)\,dz
\label{eqaccoppiateug1}
\ee
where
\be K(t,z;\kappa) = \frac{\kappa}{\pi} \left(\dfrac1{(z-t)^2 + \kappa^2}
+ \dfrac1{(z + t)^2 + \kappa^2}\right)\,;\qquad \text{with:}\;\kappa = \ell/a\,.
\label{kernelsing}\ee
$a$ is the disk radius and $\ell$ the distance between disks.
Clearly the solutions depend parametrically on $\kappa$, so a more correct notation would be $F_i(t;\kappa)$, but we omit the second argument when it is not necessary.
The charges on the disks are
\be Q_1 = a\,\frac2\pi \int_0^1 F_1(t)\,dt \,, \qquad Q_2 = a\,\frac2\pi \int_0^1 F_2(t)\,dt\,.\label{cariche1}\ee
Specializing \eqref{eqaccoppiateug1} to the case $V_1 = V_0 , V_2 = 0$ and with $F_i = V_0 f_i$ we have
\be
1 = f_1(t) +\int_0^1 K(t,z;\kappa) f_2(z)\,dz\,;\quad 
0 = f_2(t) +  \int_0^1 K(t,z;\kappa) f_1(z)\,dz
\label{eqaccoppiateug1.2}
\ee
and from \eqref{cariche1} it follows
\be C_{11} = a\,\frac2\pi \int_0^1 f_1(t)\,dt \,, \qquad C_{12} = a\,\frac2\pi \int_0^1 f_2(t)\,dt\,.\label{cariche1.3}\ee
In the case $V_1 = V_0 = - V_2$ and with $F_i = V_0 \tilde f_i$
\be
1 = \tilde f_1(t) +\int_0^1 K(t,z;\kappa) \tilde f_2(z)\,dz\,;\quad 
-1 = \tilde f_2(t) +  \int_0^1 K(t,z;\kappa) \tilde f_1(z)\,dz\,.
\label{eqaccoppiateug1.2love}\ee
It is evident that $\tilde f_1 = f_L = -\tilde f_2$, then by solving the single equation (the original Love equation):
\be
1 =  f_L(t) - \int_0^1 K(t,z;\kappa)  f_L(z)\,dz\,.\label{eqaccoppiateug1.3love}\ee
we solve the problem of two disks at opposite potential. Remembering that in this case the potential difference is $2$ we have
\be C = \frac{Q}{\Delta V} = \frac12 \frac{2a}{\pi}\int_0^1 f_L(t) dt \equiv \frac{2a}{\pi}\int_0^1 f_H(t) dt\,. \label{eqaccoppiateug1.4love}\ee
For uniformity it is convenient to define $f_H = f_L/2$, in such a way that the normalizations in all the integrals are the same. Clearly $f_H$
satisfy the integral equation \eqref{eqaccoppiateug1.3love} with $1/2$ on the l.h.s..

It is quite difficult to extract the asymptotic short distance behavior, $\kappa\to 0$, from \eqref{eqaccoppiateug1.3love}, mainly because at $\kappa\to 0$ the ``lorentzian'' kernel tends to a delta function and the equation becomes singular. This difficult task has been performed by 
V. Hutson\cite{Hutson} who obtained
\be f_H(t;\kappa) = \frac12 \left\{ \frac1\kappa \sqrt{1-t^2} + (1-t^2)^{-1/2}\frac1{2\pi}\left(
1 + \log\frac{16\pi}{\kappa} - t \log\frac{1+t}{1-t}\right)\right\} + {o}(1)\,.\label{hutson}
\ee
Terms going to $0$ as $\kappa\to 0$ are neglected: in the following we will use $f\sim g$ if $f = g + o(1)$. 
The estimate \eqref{hutson} is valid away from the edge $t \sim 1$, but this is sufficient for the computation of capacity, because an interval of order $\kappa$ gives negligible contributions, as it is easily shown that the solution is bounded on the whole interval, and in effect is of order 1 at $t\sim 1$. Integration of \eqref{hutson} gives directly the classical Kirchhoff\cite{Kirchh} formula for the relative capacity $C$
\be
C_K(\ell) = \frac{a^2}{4 \ell} + \frac{a}{4\pi}\left[ \log\left(16\pi \frac a \ell\right) - 1\right]\,.
\label{parametri1.5}\ee 
We remember that for two equal conductors the relative capacity is related to coefficients $C_{ij}$ by
\be C = \dfrac{C_{11} - C_{12}}{2}\,. \label{defc}\ee
Let us note that the divergent behavior 
of the solution for $\kappa\to 0$ is specifically due to the fact that the kernel $K$ becomes a $\delta$ in this limit: this rules out evidently any finite solution
in the limit $\kappa\to 0$. Translating \eqref{eqaccoppiateug1.3love} in matrices language the solution is
\[ f_L = (1 -  K)^{-1} u \]
where $u$ is a vector filled with 1's. When the operator $K$ has an eigenvalue 1 the solution diverges.

For two equal disks of radius $a$ the limit $\kappa\to0$ produces a single disk with the same radius and with a known capacitance $C_F = 2a/\pi$.
The arguments introduced in section \ref{constraints} and the expression eq.\eqref{limitiC2}, fix in this case the asymptotic behavior of $C_{11}$ and $C_{12}$:
\begin{subequations}\label{cijdischiuguali}
\begin{align}
C_{11} &\simeq C_{K}(\ell) +\frac a{2\pi} =
\frac{a^2}{4 \ell} + \frac{a}{4\pi}\left[ \log\left(16\pi \frac a \ell\right) + 1\right]\,;\\
-C_{12} & \simeq C_{K}(\ell) - \frac a{2\pi} = 
\frac{a^2}{4 \ell} + \frac{a}{4\pi}\left[ \log\left(16\pi \frac a \ell\right) -3 \right]\,.
\end{align}
\end{subequations}

To confirm our statement \eqref{cijdischiuguali} we have to solve \eqref{eqaccoppiateug1.2} and
find the expected result from \eqref{cariche1.3}. 
The equations can be decoupled using the change of variables:
\[ f_1(t) = \frac12 f(t) + \frac12 \delta f(t)\,;\qquad f_2(t) = -\frac12 f(t) + \frac12 \delta f(t)\]
Adding and subtracting the two equations \eqref{eqaccoppiateug1.2} we obtain:
\begin{subequations}\label{dec1}
\begin{align}
1 &= \delta f(t) + \int_0^1 K(t,z;\kappa) \delta f(z)\,dz\label{dec1a}\\
1 &= f(t) - \int_0^1 K(t,z;\kappa) f(z)\, dz \label{dec1b}
\end{align}
\end{subequations}
The last equation is the usual Love equation, \eqref{eqaccoppiateug1.3love}, then $f(t) = f_L(t)$. Eq.\eqref{dec1a} is similar to
\eqref{eqaccoppiateug1.3love} but with a {\em crucial} difference in sign: now the operator $(1+K)$ is clearly invertible  for $\kappa\to 0$ and the solution of \eqref{dec1a} in this limit, where $K$ behaves like an identity operator, is
\be \delta f(t) = \frac12 \label{valoredf}\ee
Collecting the two results together we have for $\kappa\to 0$
\be f_1(t) \sim \frac12 f_L(t) +\frac14 \equiv f_H(t) + \frac14 \,;\qquad f_2(t) \sim -\frac12 f_L(t) +\frac14
\equiv - f_H(t) +\frac14
\label{df1234}\ee
Integrating and using \eqref{eqaccoppiateug1.4love} we have finally
\begin{align}
C_{11} = \frac{2a}{\pi}\int_0^1 f_1(t) dt = C_K + \frac{a}{2\pi}\,;\qquad
- C_{12} = - \int_0^1 f_2(t) dt = C_K - \frac{a}{2\pi}
\end{align}
i.e. \eqref{cijdischiuguali}.

From the physical point of view it is instructive to expose the mechanism of cancellation of divergences in \eqref{cijdischiuguali}
by translating in equations the simple considerations of the foregoing section.
The physical procedure outlined there is here realized by considering the equations \eqref{eqaccoppiateug1} with $V_1 = V_2 = 1$
and $F_i = V_i f_i = f_i$.
The two equations are identical and we arrive at the single relation
\be 1 = f_1(t) + \int_0^1 ds\, K(t,s,\kappa) f_1(s)\label{eqv1v1} \ee
This is identical to \eqref{dec1a} and has the same asymptotic solution for $\kappa\to 0$, i.e. $f_1(t)\sim 1/2$.
From the general relations \eqref{eqaccoppiateug1} and \eqref{cariche1} the integral of the solution in this case gives $C_{11}+ C_{12}$, i.e.
as $\kappa\to 0$:
\be C_{11} + C_{12} = \frac{2a}\pi \int_0^1 f_1(t) dt \to \frac{2a}\pi \frac12 = \frac{a}{\pi} = \frac{C_1}{2}\,,\label{valoreC1}\ee
which is the correct result. The  eq.\eqref{valoreC1} is a {\em computation} of the capacity of a single disk, independent in principle from the usual one\cite{LL}. A similar observation will be discussed for the general case of large distance expansion in the next section.

\section{Coaxial disks with different radii\label{sezdiscussione}}

Let us consider now the more general case of two coaxial disks with radii $a, c$ with $b = c/a > 1$. The distance between the disks is $\ell$.
In this section we summarize and discuss the results obtained in this case, the formal developments are given in the next two sections.

\subsection{Equations and potential for different disks}
The first result is that we 
can deal with this problem using the previous approach, based on the separability of Laplace equation in cylindrical coordinates. One arrives at a system of equations similar to the previous one: 
\be
V_1 = F_1(t) +\int_0^b K(t,z;\kappa) F_2(z)\,dz\,;\quad 
V_2 = F_2(t) +  \int_0^1 K(t,z;\kappa) F_1(z)\,dz\,.
\label{eqaccoppiatediv1}
\ee
The notation is the same of previous section: $\kappa = \ell/a$ and the kernel $K$ is always given by \eqref{kernelsing}.
Charges are given by
\be Q_1 = \frac{2a}{\pi} \int_0^1 F_1(t)\,dt\,;\quad Q_2 = \frac{2a}{\pi} \int_0^b F_2(t)\,dt\,.\label{caichedef}\ee
The exact form of the potential is
\be
\phi(r,z) = \frac{2a}{\pi}\text{Re}\left[
\int_0^1 dt \dfrac{F_1(t)}{\sqrt{r^2 + (|z|- i a t)^2}} + 
\int_0^b dt \dfrac{F_2(t)}{\sqrt{r^2 + (|z - \ell|- i a t)^2}}\right]\,.\label{potenziale}
\ee
where $r^2 = x^2+y^2$. A proof of these three statements is given in section~\ref{provaequazione}.
\subsection{Matrix of Cij in the near limit}
A second result is that the system \eqref{eqaccoppiatediv1} can be solved, in the asymptotic regime $\kappa\to 0$, in a region excluding an interval of order $\kappa$ near the boundary $t= 1$. This is sufficient to compute the capacitance coefficients extending the classical Kirchhoff formula to this more general case. The result is
\be C_{11} = C_{11}^{(\infty)} + \delta_{11}(b)\,;\quad C_{12} = - C_{11}^{(\infty)} + \delta_{12}(b)\,;
\quad C_{22} = C_{11}^{(\infty)} + \delta_{22}(b) \label{valoriCij}\ee
where 
\be C_{11}^{(\infty)} = \frac{2 a}{\pi}\int_0^1 f_L(t;2\kappa) \sim \frac{a}{4\kappa} + \frac{a}{2\pi}\left[ \log\bigl(\frac{8\pi}\kappa\bigr) - 1\right] \label{c11inf} 
\ee
is the value of the self-induction coefficient for $b\to\infty$, as will be proven below, and
\begin{subequations}\label{valoriCijbis}
\begin{align}
\delta_{11}(b) &= \frac{a}{\pi}\left[ b - \sqrt{b^2-1} - \frac12 {\rm arctanh}\frac1b\right]
\mathop{\sim}_{b \gg 1} - \frac{a}{24\pi b^3}
\\
\delta_{12}(b) &= \frac{a}{2\pi} {\rm arctanh}\frac1b\,
\mathop{\sim}_{b \gg 1} \frac{a}{2\pi b}
;\qquad \delta_{22}(b) = \frac{2a}\pi\,b - \delta_{11}(b) - 2 \delta_{12}(b) \mathop{\sim}_{b \gg 1} \frac{2a}\pi\,b - \frac{a}{\pi b}\,.
\end{align}
\end{subequations}
$f_L(t;2\kappa)$ is the solution of the Love's equation with scale $2\kappa$.
It is also convenient to put these relations in the form:
\be
C_{11}\sim - C_{12} +\frac{a}{\pi}\left(b - \sqrt{b^2-1}\right)\,;\quad C_{22}\sim - C_{12} + \frac{a}{\pi}\left(b + \sqrt{b^2-1}\right)\,.
\label{cvsc12}\ee
The proof of these statements is in section \ref{risasintotici}, where the asymptotic form of the solutions is also computed.

\subsection{Discussion of the asymptotic solutions}

Before starting the mathematical analysis of the results may be useful to have a look at the solution and perform some checks.

Let us start from large distances, $\kappa\gg 1$.
As in the case of equal disks it is quite easy to make an expansion in powers of $1/\kappa$: the solutions $F_i$ are polynomial in $t$, as can be easily seen by an iterative solution of \eqref{eqaccoppiatediv1}. This allows us to perform a check of some historical interest: computing the solutions and integrating we obtain the capacitance matrix
\begin{subequations}\label{coeffC}
\begin{align}
C_{11} &= \frac{2 a}{\pi}
\bigl( 1+ \frac{4 a {c}}{\pi ^2 \ell^2}
-\frac{8 a {c} \left(\pi ^2 a^2-6 a {c}+\pi ^2 {c}^2\right)}{3 \pi ^4
   \ell^4} +\ldots \bigr)\\
   C_{12} &= -\frac{4 a{c}}{\pi^2}\frac1\ell +
   \frac{4 a {c} \left(\pi ^2 a^2-12 a {c}+\pi ^2 {c}^2\right)}{3 \pi ^4}\frac{1}{\ell^3}+\ldots = C_{21}\\
 C_{22} &= \frac{2{c}}{\pi}
\bigl( 1+ \frac{4 a {c}}{\pi ^2 \ell^2}
-\frac{8 a {c} \left(\pi ^2 a^2-6 a {c}+\pi ^2 {c}^2\right)}{3 \pi ^4
   \ell^4} +\ldots \bigr)\end{align}
   \end{subequations}
Inverting the matrix we can write the potential coefficients $M_{ij}$ and we obtain, in usual units:
\begin{subequations}\label{coeffM}
\begin{align}
M_{11} &= \frac{\pi}{2a} - \frac8{45\pi}\frac{{c}^5}{\ell^6} +
\frac{32 c^5}{15\pi}\bigl(\frac{ a^2}{3}+\frac{ {c}^2}{7  }\bigr)\frac1{\ell^8}-
\frac{16 c^5}{\pi}\bigl(\frac{ a^4}{9 }+\frac{2 a^2 {c}^2}{15 }+\frac{13 {c}^4}{525
    }\bigr)\frac{1}{\ell^{10}}\nonumber \\
   & + 
   \frac{32 {c}^5}{\pi}  \left(\frac19 a^6+\frac9{35} a^4 {c}^2+\frac{127}{945} a^2 {c}^4+\frac{151}{10395} {c}^6\right)\,\frac{1}{\ell^{12}}
   +\ldots\\
M_{22} &= M_{11} (a \leftrightarrow {c} )\nonumber\\
M_{12} &= \frac1\ell - \frac13(a^2+{c}^2)\frac{1}{\ell^3}
+\bigl(
\frac{a^4}{5}+\frac{2 a^2 {c}^2}{3}+\frac{{c}^4}{5}\bigr)\frac{1}{\ell^5}-
\bigl(
\frac{a^6}{7}+a^4 {c}^2+a^2 {c}^4+\frac{{c}^6}{7}
\bigr)\frac1{\ell^7}\nonumber\\
& + \bigl(
\frac{a^8}{9}+\frac{4 a^6 {c}^2}{3}+\frac{14 a^4 {c}^4}{5}+\frac{4 a^2
   {c}^6}{3}+\frac{{c}^8}{9}
\bigr)\frac1{\ell^9}\nonumber\\
&- \bigl(\frac{a^{10}}{11} + \frac53 a^8 c^2 + 6 a^6 c^4 + 6 a^4 c^6 + \frac53 a^2 c^8 +\frac{c^{10}}{11} 
-\frac{128}{675\,\pi^2} a^5 c^5\bigr)\,\frac1{\ell^{11}}+\ldots
\end{align}
\end{subequations}
The coefficient $M_{11}, M_{22}$ up to order $1/\ell^{10}$ and $M_{12}$ up to order $1/\ell^7$ have been computed by Maxwell in ref.\cite{maxw2} and 
coincide 
with expressions \eqref{coeffM}.
The first terms in \eqref{coeffC}, up to order $1/\ell^4$ included, are in agreement with the general result of ref.\cite{mac}, expressed in terms of the self capacitance, the polarizability tensor $\alpha_{ij}$ and the quadrupole moment $D_{ij}$ of the conductors. In the case at hand for a disk of radius $a$ the relevant parameters are\cite{LL}
\be C_1 = \frac{2a}{\pi}\,;\qquad \alpha_{zz} = 0\,;\quad D_{zz} = -\frac23 a^2\,. \label{valorec1alpha}\ee
Again as in the previous section we underline that we presented here \eqref{valorec1alpha} as a check of the procedure, but in effect it is also an alternative method to compute the listed parameters for a disk.

The second region of interest is $\kappa\to 0$.
In this limit the two conductors form a disk of radius $c = a b$ (the bigger of the two) so we expect from \eqref{limitiCa}
\be \lim_{\kappa\to 0} \Bigl( C_{11} + 2 C_{12} + C_{22}\Bigr) = 2 \frac{c}\pi = \frac{2 a}{\pi}\, b  \label{limit1}\ee
and
\be \lim_{\kappa\to 0} \Bigl( C_{11} +  C_{12}\Bigr) < \infty \,;\qquad \lim_{\kappa\to 0} \Bigl( C_{22} +  C_{12}\Bigr)
< \infty \,. \label{limit2}\ee
These expectations are confirmed by \eqref{valoriCij}, where also the value of separated limits is given.

Let us note that the definition of short distance in the present case, while clear from a mathematical viewpoint, require some physical specifications.
We have {\em two} possibly small parameters, $\ell$, the distance between the disks, and, in some cases, $c-a$, or, in scaled units, $\kappa$ and $b-1$. On the mathematical side the limit is understood in the sense $\kappa\to 0$ with $b>1$ fixed. In the following it will be clear that the limits $b\to 1$ and $\kappa\to 0$ do not commute so a special care must be taken when comparing capacitance with approximate numerical results 
in this regime.

\subsection{Physical and numeric analysis of results.}

Having passed these preliminary checks let us now sketch the procedure followed in the analysis of the problem.
A clue to the solution of problem \eqref{eqaccoppiatediv1} comes from the physical considerations given in the section \ref{constraints}.
We expect that in the $\kappa\to 0$ region the complex of force lines is determined mainly by the smaller disk, then let us first
consider the limit case $V_1=1, V_2=0$ and $b\to \infty$. The large disk is placed, to fix the ideas, in the $x-y$ plane 
and grounded, i.e. $V_2=0$,
the smaller disk is placed
at $z= a \kappa = \ell$.

Using the image method for the solution it is clear that in the half-space $z>0$ the potential is identical to the one obtained by two disks, oppositely charged, at distance $2\ell$, the second disk being the image of the first at coordinate $z=-\ell$. So we are dealing with a problem of two identical disks, and the charge accumulated on disk 1, i.e. $C_{11}$, will be simply given by the usual expression
for two disks with a potential difference $\Delta V=2$ at distance $ \ell = 2 \kappa a$, i.e. (see eq.\eqref{parametri1.5})
\be C_{11} \sim 2 C_K(2 \kappa a) = 
\frac{a}{4\kappa} + \frac{a}{2\pi}\left[ \log\bigl(\frac{8\pi}\kappa\bigr) - 1\right] \equiv C_{11}^{(\infty)}
\label{c11blarge}\ee
We can learn two lessons from this result
\begin{itemize}
\item[1)] The ``geometrical'' term in the capacitance is dictated by the smallest disk, as expected from elementary considerations on the flux of the electric field.
\item[2)] The logarithmic (and finite) part of edge corrections is {\em different} from \eqref{cijdischiuguali}, this signals a crossover region
in the limit $\kappa\to 0 $ when $b\to 1$, i.e. the two limits do not commute.
\end{itemize}
In this limit the second disk (the large one) has clearly the opposite charge of disk 1, i.e. $C_{12}= - C_{11}$, a result expected 
by simple considerations on flux lines, \label{c12ugc11} in the limit $b\to \infty$.

The same conclusions can be drawn more formally from the equations \eqref{eqaccoppiatediv1}, which for the normalized functions read
\be 1 = f_1(t) + \int_0^b K(t,s;\kappa) f_2(s)\,ds \,;\qquad 0 = f_2(t) + \int_0^1 K(t,s;\kappa) f_1(s)\,ds
\label{f1f2}\ee
In the limit $b\to \infty$ substituting the second equation in the first and performing the intermediate integral we have
\be 1 = f_1(t) - \int_0^1 \dfrac{2\kappa}{\pi}\left[\dfrac{1}{4\kappa^2+ (t-s)^2} + \dfrac{1}{4\kappa^2+ (t+s)^2}\right]\,f_1(s)\,ds
\equiv f_1(t) - \int_0^1 K(t,s;2\kappa) f_1(s) \label{f1f2.1}\ee
i.e. the Love equation with a scale doubled. Using the definition already introduced $\frac12 f_L = f_H$ we write the solution at once
\be f_1(t) = 2 f_H(t;2\kappa) 
\sim 
 \frac1{2\kappa} \sqrt{1-t^2} + (1-t^2)^{-1/2}\frac1{2\pi}\left(
1 + \log(\frac{8\pi}{\kappa}) - t \log\frac{1+t}{1-t}\right)
\label{f1f2.2}\ee
which by integration gives \eqref{c11blarge}.

This simple result is the one on which is built the solution in section \ref{risasintotici}.

As the following formal analysis can be rather tedious maybe is of some interest for the reader to have a look at the results 
in a particular case
from a graphical point of view. To illustrate the point we take the case $b=1.1$, which {\it a priori} could be problematic as $b$ is not so large.

\begin{figure}[!ht]
\begin{center}
\includegraphics[width=0.85\textwidth]{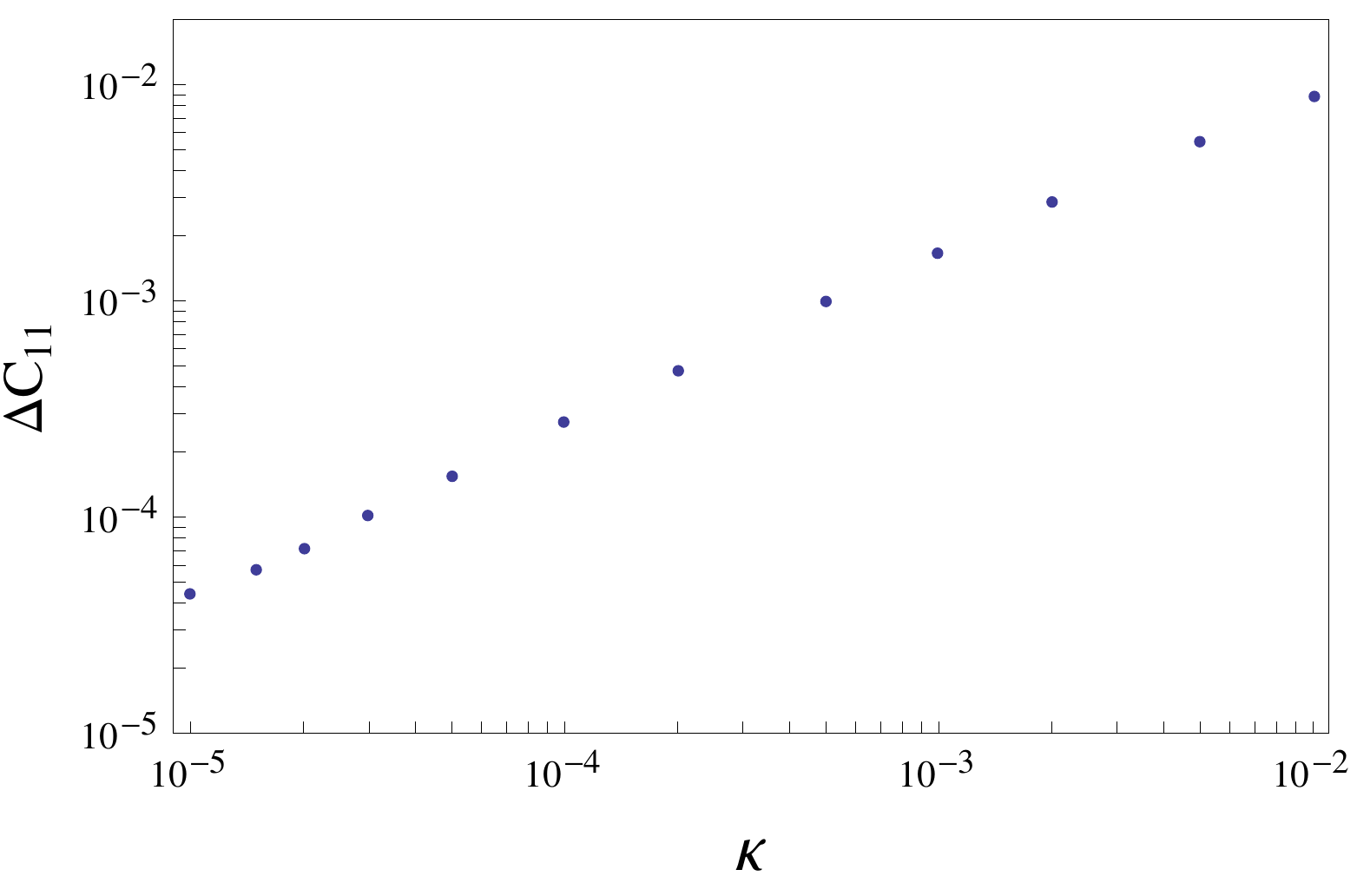}
\caption{Difference between computed and predicted asymptotic values for $C_{11}$ (circles) for $b=1.1$\label{fig1a}}
\end{center}
\end{figure}

\begin{figure}[!ht]
\centering\includegraphics[width=0.85\textwidth]{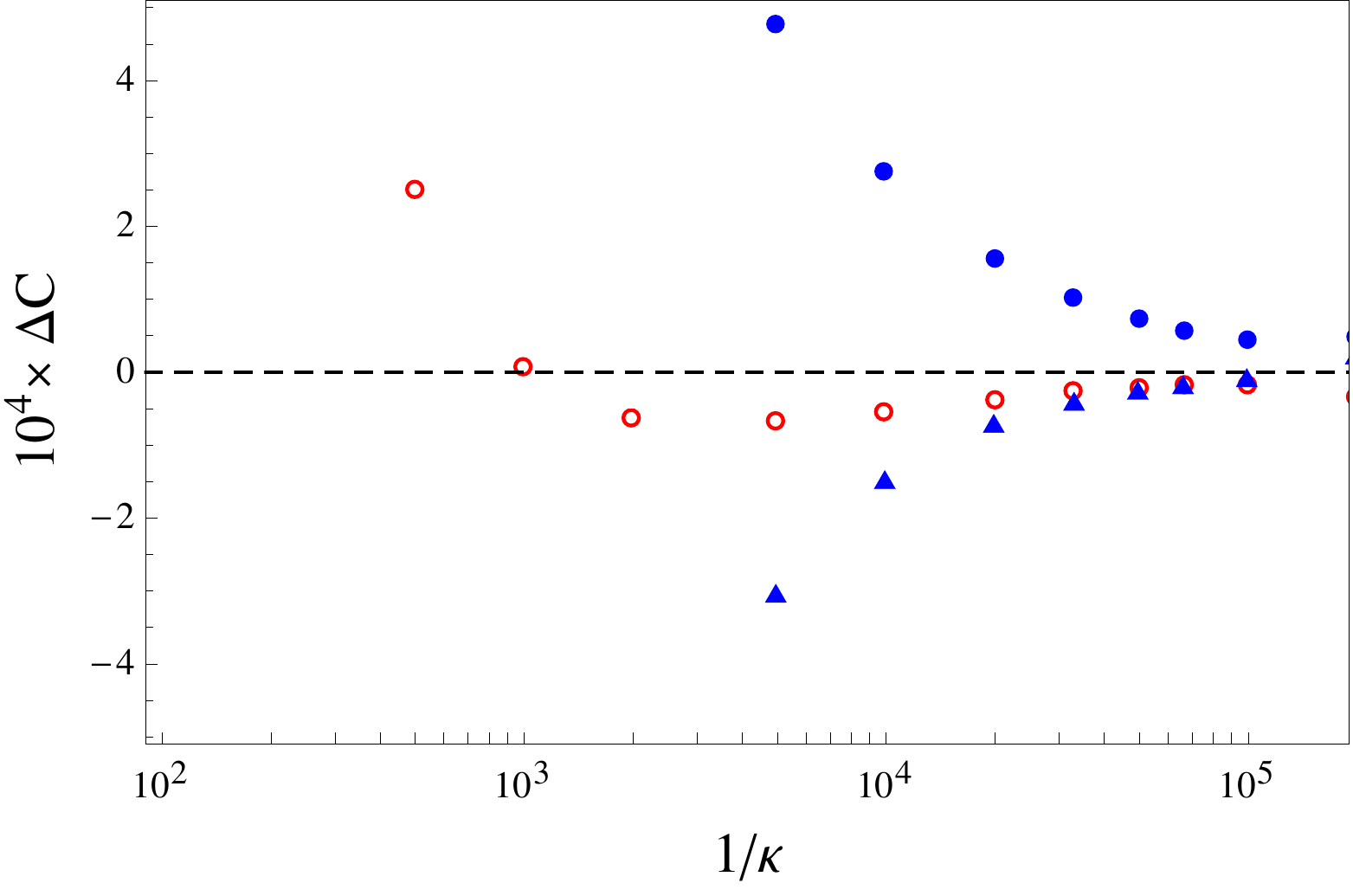}
\caption{Differences between computed and predicted asymptotic values for $C_{11}$ (full circles), $C_{22}$ (triangles) and $C_{12}$ (empty points) for $b=1.1$}
\label{fig1b}
\end{figure}
%\vspace*{-5pt}

In figure \ref{fig1a} the difference between the numerical solution of system \eqref{eqaccoppiatediv1} and the asymptotic prediction
\eqref{valoriCij} is plotted. One can appreciate that the agreement is quite good.
In figure \ref{fig1b} is reported the same kind of differences on a linear scale for the three coefficients. The non monotonic
approach to the limit for $C_{12}$ is characteristic in the case of $b\sim 1$ and disappears for large $b$.
The details of numerical solutions of system \eqref{eqaccoppiatediv1} will be given elsewhere.
Let us note that in the range $\kappa\sim 10^{-4}-10^{-5}$ the absolute value of capacitance coefficients is of order $10^4-10^5$, so the agreement
shown in the figure is, on an absolute scale, of the order of one part in $10^7-10^8$. For the analogous case of identical disks at least four order of magnitudes are simply due to the geometrical capacitance, so the agreement would be good but not so impressive. For {\em different} disks
this agreement is a quite strong numerical evidence for the picture sketched in sect.\ref{constraints}: only the smallest disk dictates the divergences.

\begin{figure}[!ht]
\centering\includegraphics[width=0.49\textwidth]{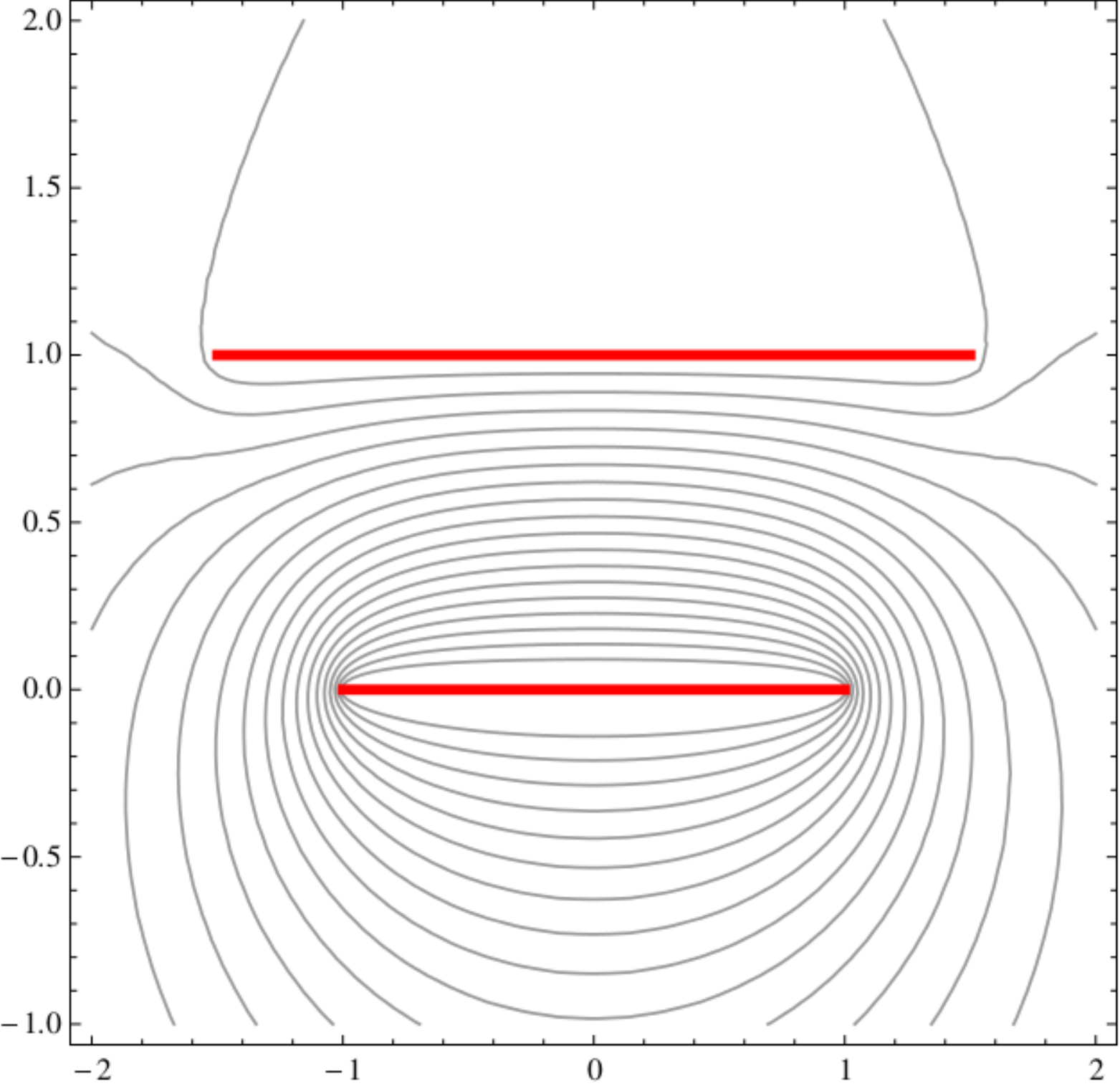}
\includegraphics[width=0.49\textwidth]{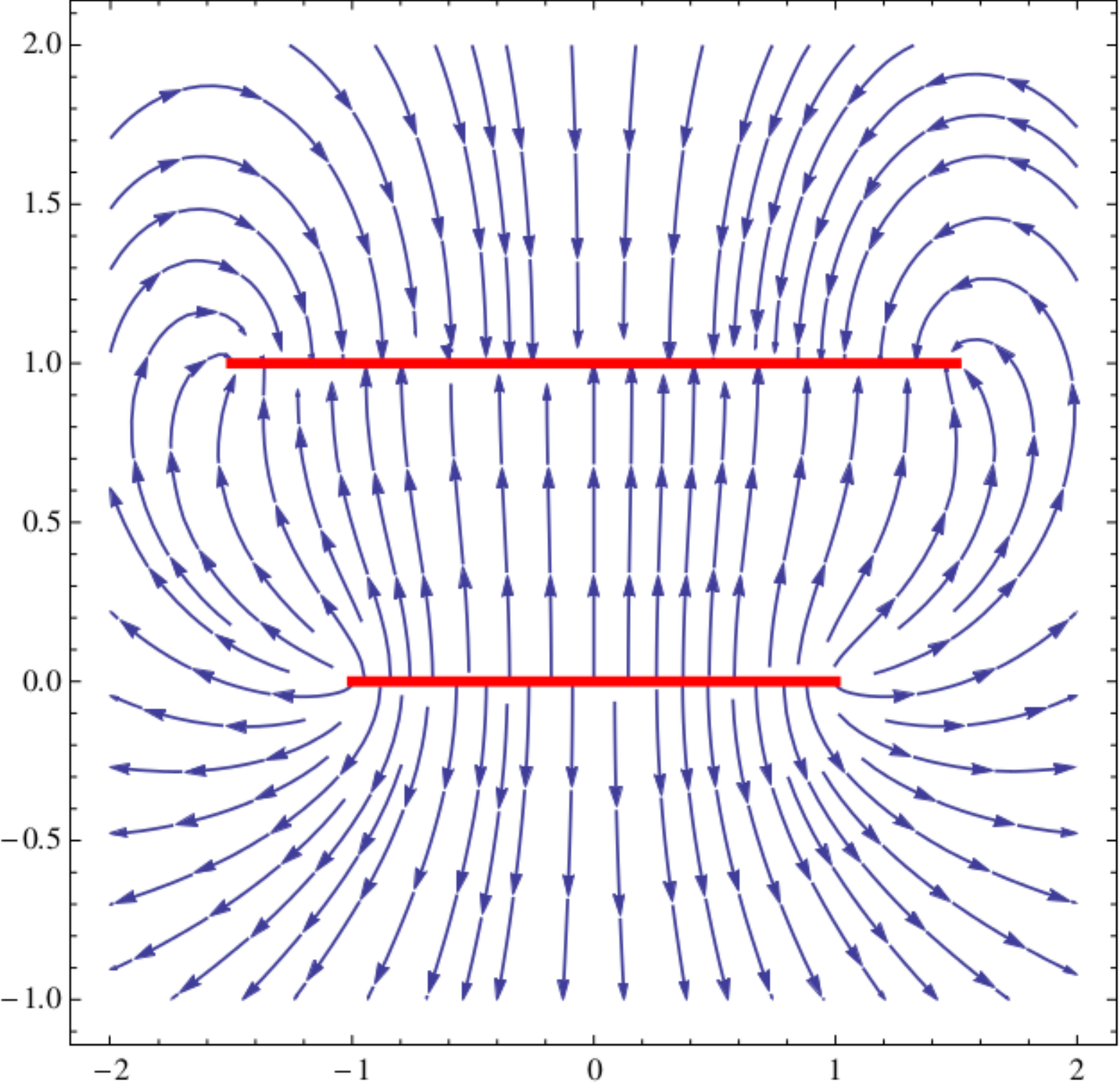}
\caption{Equipotential lines and electric field stream lines for $b=1.5$ and $\ell = 1$. Smaller disk is at potential $V_1=1$, larger disk at potential $V_2=0$.
Equipotential lines are in the range $0.05-0.9$, separated by $0.05$. The density of streamlines is unrelated to the value of the electric field..}
\label{fig3a}
\end{figure}

Finally, in fig.\ref{fig3a} we give a plot of equipotential lines and flux lines computed from \eqref{potenziale}.

\section{The integral equation\label{provaequazione}}
Let us consider two coaxial disks of radii $a$ and $c$, with $b = c/a > 1$. The first is placed on the plane $z=0$ and centered at the origin
of the reference frame. The second disk is in the plane $z=\ell$. The potentials are respectively $V_1$ and $V_2$. Laplace equation is clearly separable in cylindrical coordinates and it is easily seen that the general solution vanishing at infinity has the form
\be \phi( r,z) = \int_0^\infty \left[ G_1(q) e^{-|z|q}  + G_2(q) e^{-|z-\ell|q}\right] \frac1q J_0(q r) \,dq
\label{equaz1.1}\ee
where $ r^2 = x^2+y^2$ and $J_0$ is the Bessel function
of order 0. In this section we freely use a certain number of nontrivial integrals
of Bessel functions, the reader is referred to ref.\cite{sne,Carlson} for further information on this subject.

The first boundary condition is on the potentials:
\begin{subequations}\label{vincoli1}
\begin{align}
V_1 &= \int_0^\infty \left[ G_1(q)  + G_2(q) e^{-\ell q}\right] \frac1q J_0(q r) \,dq\,;\qquad  r < a\\[6pt]
V_2 & = \int_0^\infty \left[ G_1(q)e^{-\ell q}  + G_2(q)\right] \frac1q J_0(q r) \,dq\,;\qquad  r < b a
\end{align}
\end{subequations}
The second condition comes from the continuity of electric field in the region outside the conductors. From \eqref{equaz1.1} appears that a possible discontinuity arises in the $z$ derivative of $\phi$. The $z$ derivative of $\phi$ is easily computed and imposing to it the continuity at the planes  $z = 0$
and $z = \ell$ outside the conductors fixes respectively the conditions

\begin{subequations}\label{vincoli2}
\begin{align}
\Delta E_z^{(1)} &= \int_0^\infty G_1(q)  J_0(q r) \,dq\,= 0;\qquad  r > a\\[6pt]
\Delta E_z^{(2)} & = \int_0^\infty  G_2(q)  J_0(q r) \,dq\,= 0;\qquad  r > b a\,.
\end{align}
\end{subequations}
The solution of the Laplace equation with boundary conditions \eqref{vincoli1} and \eqref{vincoli2} fixes the potential $\phi$. The discontinuity of the subsequent computed electric field on the disks gives the charge density
\begin{subequations}\label{vincoli3}
\begin{align}
\sigma_1(r)  &= \frac{1}{2\pi}\int_0^\infty G_1(q)  J_0(q r) \,dq\,;\qquad  r < a\\[6pt]
\sigma_2(r)  & = \frac{1}{2\pi}\int_0^\infty  G_2(q)  J_0(q r) \,dq\,;\qquad  r < b a
\end{align}
\end{subequations}
It is convenient to define adimensional variables in the form
\be \ell = \kappa\,a \,;\quad  r = x a\,;\quad q = u/a \,;\qquad G_1(q) \equiv g_1(u)\,;\quad G_2(q)\equiv g_2(u)\label{variabiliadimensionali}\ee
Using these variables the equations \eqref{vincoli1} and \eqref{vincoli2} take the form
\begin{subequations}\label{vincoli1bis}
\begin{align}
V_1 &= \int_0^\infty \left[ g_1(u)  + g_2(u) e^{-\kappa u }\right] \frac1u J_0(u x) \,du\,;\qquad x < 1 \label{vincoli1bisa}\\[6pt]
V_2 & = \int_0^\infty \left[ g_1(u)e^{-\kappa u}  + g_2(u)\right] \frac1q J_0(u x) \,du\,;\qquad x < b \label{vincoli1bisb}
\end{align}
\end{subequations}
and
\begin{subequations}\label{vincoli2bis}
\begin{align}
\Delta E^{(1)}_z(x) &= \int_0^\infty g_1(u) J_0(u x) \,du\,= 0;\qquad x > 1\\[6pt]
\Delta E^{(2)}_z(x) & = \int_0^\infty  g_2(u)  J_0(u x) \,du\,= 0 ;\qquad x > b 
\end{align}
\end{subequations}
while the charges on the conductors, obtained by integrating $\sigma_1, \sigma_2$ in $2\pi  r\,d r$ are
\be
Q_1 =  a \int_0^1 \!\!d x\,x\,   \int_0^\infty\hskip-5pt du\; g_1(u)  J_0(u x) \,;\quad
Q_2 = a \int_0^{b}\!\!d x \,x\,\int_0^\infty\hskip-5pt du\;  g_2(u)  J_0(u x ) \label{vincoli4bis}\ee
The next step is an integral transformation for the functions $g_1, g_2$:
\begin{subequations}\label{sostituzionegF}
\begin{align}
g_1(u) &= 2\dfrac{u}{\pi} \int_0^1 F_1(t) \cos(u t)\,dt \\[6pt]
g_2(u) &=  2\dfrac{u}{\pi} \int_0^b F_2(t) \cos(u t)\,dt
\end{align}
\end{subequations}
This is the point where the problem of two disks differ from the case with equal disks. It turns out that the simple
change in the range of $t$ is sufficient to take care of the difference in the size of disks.
Substitution in \eqref{vincoli2bis} gives rise to expressions
\begin{align*}
\Delta E^{(1)}_z(x) = &\frac2\pi\int_0^1\! dt F_1(t) \int_0^\infty u \cos(u t) J_0(u x) \,du =
\frac2\pi\int_0^1 dt F_1(t) \frac d{dt} \int_0^\infty \sin(u t) J_0(u x) \,du
\\[6pt] 
\Delta E^{(2)}_z(x) = &\frac2\pi\! \int_0^b F_2(t) \int_0^\infty  u  \cos(u t)  J_0(u x) \,du =
\frac2\pi \int_0^b\! dt F_2(t) \frac{d}{dt}\int_0^\infty  \sin(u t)  J_0(u x) \,du 
\end{align*}
The integral of the Bessel function with trigonometric functions are known:
\be 
\int_0^{\infty } J_0(u x) \cos (t u) \, du=\frac{\theta (x-t)}{\sqrt{x^2-t^2}}\,;\quad
\int_0^{\infty } J_0(u x) \sin (t u) \, du=\frac{\theta (t-x)}{\sqrt{t^2-x^2}}\label{intebralij0b}\ee
and as the integrals in $t$ for $\Delta E^{(1)}$ and $\Delta E^{(2)}$ extend to 1 and $b$ respectively, the $\theta$ function assures that
boundary conditions \eqref{vincoli2bis} are satisfied.

Substitution of \eqref{sostituzionegF} in \eqref{vincoli1bis} gives the integral equations \eqref{eqaccoppiatediv1}.

The proof below parallels step by step the derivation given in ref.\cite{Carlson}.
Let us consider \eqref{vincoli1bisa}, the substitution produces
\begin{align*}
\frac\pi2 V_1 
&= \int_0^1 dt\int_0^\infty\hskip-5pt du\;  F_1(t) \cos(u t) J_0(u x)  + \int_0^b dt \int_0^\infty\!\! du\;  F_2(t) \cos(u t) e^{-\kappa u}J_0(u x) 
\end{align*}
and using the first integral in \eqref{intebralij0b}
\be
\frac\pi2 V_1 = \int_0^x dt \dfrac{F_1(t)}{\sqrt{x^2-t^2}} +
 \int_0^b dt\int_0^\infty\hskip-5pt du\;  F_2(t) \cos( u t) e^{-\kappa u} J_0(u x)\label{proofeq1.1}
\ee
This is an Abel integral equation for $F_1$, of the general form
\be
G(x) = \int_0^x \frac{F(t)}{\sqrt{x^2-t^2}} \, dt\,;\quad \Rightarrow \quad
F(t) = \frac{2}{\pi} \frac{d}{dt}
\int_0^t \dfrac{ x\, G(x)}{\sqrt{t^2-x^2}} dx 
\label{abel1}\ee
In our case
\[ G(x) = \frac\pi2 V_1 - \int_0^b dz\int_0^\infty\!\!du\;  F_2(z) \cos(u z) e^{-\kappa u}J_0(u x) \]
then from \eqref{abel1}, performing an elementary integral:
\[
F_1(t) = V_1 - \frac2\pi\int_0^b dz F_2(z) \frac{d}{dt}\int_0^t dx\;
\left[ \int_0^\infty\!\!du\;\; e^{-\kappa u} \cos(u z) J_0(u x) \dfrac{x}{\sqrt{t^2-x^2}}\right]\]
Interchanging the order of the integrals and computing the integral in $x$
\be \int_0^t dx J_0(u x) \dfrac{x}{\sqrt{t^2-x^2}} = \dfrac{\sin(tu)}{u} \label{integrale1}\ee
the derivative with respect to $t$ gives:
\be F_1(t) = V_1 - \frac2\pi \int_0^b dz F_2(z) \int_0^\infty \!\!du\; e^{-\kappa u} \cos(u z) \cos(t u)
\label{equaz2.2}
\ee
The last integral is elementary
\be \int_0^\infty\hskip-5pt du\; e^{-\kappa u} \cos(z u) \cos(t u) = \frac{\kappa}{2} \left(\dfrac1{(z-t)^2 + \kappa^2}
+ \dfrac1{(z+t)^2 + \kappa^2}\right)\label{integrale4}\ee
and we finally obtain
\begin{align}
F_1(t) &= V_1 - \int_0^b K(t,z) F_2(z) \;dz\\
K(t,z) &= \frac{\kappa}{\pi} \left(\dfrac1{(z-t)^2 + \kappa^2}
+ \dfrac1{(z + t)^2 + \kappa^2}\right)
\end{align}
which is the first of our equations \eqref{eqaccoppiatediv1}. The derivation of the second equation \eqref{eqaccoppiatediv1} is almost identical and we omit it for brevity.

Now we show that the charges of the conductors are directly related to the integral of the solution $F_1, F_2$ of the system of integral equations, as anticipated in equations \eqref{caichedef} here repeated:

\be Q_1 = \frac{2a}{\pi}\int_0^1 F_1(t) dt\,;\qquad Q_2 = \frac{2a}{\pi}\int_0^b F_2(t) dt \,.\label{cariche}\ee
Let us consider $Q_2$, the charge on the larger disk. Substitution of \eqref{sostituzionegF} in \eqref{vincoli4bis} gives
\begin{align*}
Q_2 &= a \frac2\pi \int_0^b dt F_2(t) \int_0^\infty\hskip-5pt du \;u \cos(u t)\int_0^b dx\; x  J_0(u x)\\
&= a \frac2\pi \int_0^b dt F_2(t) \int_0^\infty\hskip-5pt du \;u \cos(u t) \frac{b}{u} J_1(b u)
= a \frac2\pi \int_0^b dt F_2(t) \int_0^\infty\hskip-5pt du\; b \cos(u t)  J_1(b u)\,.
\end{align*}
Using 
\[ \int_0^\infty\hskip-5pt du\; b \cos(u t)  J_1(b u) = 1\,;\qquad \text{for}\; t<b \]
one recovers \eqref{cariche}. A similar deduction can be done for $Q_1$, the charge on the smaller disk,  finding the first relation of eqs.\eqref{cariche}.

Finally let us give the explicit expression for the potential $\phi(r,z)$ in terms of the functions $F_1, F_2$.
Using the integral transformation \eqref{sostituzionegF} and the definitions \eqref{variabiliadimensionali}
we can write:
\begin{align*}
\phi(r,z) = \int_0^\infty
\frac1q dq J_0(q r)  \frac{2 a q}{\pi} \left[ \int_0^1 dt F_1(t) \cos(a q t) e^{-|z| q}+ \int_0^b dt F_2(t)\cos(a q t) e^{- |z-\ell| q}
\right]
\end{align*}
Writing $\cos(a q t) = \text{Re}[\exp(i a q t)]$, interchanging the order of the integrals  and using
the following result\cite{Carlson}
\be \int_0^\infty e^{-q y} J_0(q r) dq = \dfrac{1}{\sqrt{y^2 + r^2}}\label{integralej0sqrt}\ee
valid for $\text{Re}[y]\geq0$, we have
\be
\phi(r,z) = \frac{2a}{\pi}\text{Re}\left[
\int_0^1 dt \dfrac{F_1(t)}{\sqrt{r^2 + (|z|- i a t)^2}} + 
\int_0^b dt \dfrac{F_2(t)}{\sqrt{r^2 + (|z - \ell|- i a t)^2}}\right]\label{potenzialebis}
\ee

\section{Capacitance coefficients in the near limit\label{risasintotici}}
Using an argument based on the method of image charges, it has been shown
in section \ref{sezdiscussione}
 that for $b\to\infty$ the solution for a system of two disks with $V_1=1, V_2=0$ is expected to be 
$f_1(t) = f_L(t;2\kappa)$, where $f_L$ is the solution of the standard Love equation with a doubled scale. In this approximation $C_{11} = C_{11}^{(\infty)}= - C_{12}$, as noted at pag.\pageref{c12ugc11}. 

In this section we confirm this result and compute the corrections for finite $b$, which are shown to be finite. We also compute
the asymptotic form of the solutions $f_1(t)$ and $f_2(t)$ for different disks. There are at least three ways to obtain the 
correction to capacitance: a) compute directly the corrections to capacitance matrix, b) compute perturbatively the corrections to the basic approximation for $f_1$ given above in the system
\eqref{eqaccoppiatediv1}, c) transform the system by decoupling the variables. 

We present here only the third method, the most insightful one,
having checked that all three methods give the same result.
\subsection{Construction of the solutions for $\kappa\to 0$}
For simplicity and to refer to some known results we freely extend the solutions for $t<0$ with $f(-t) = f(t)$, this is allowed due the form of the kernel $K(t,s;\kappa)$. We will write in such a case
\[ \int_0^1 K(t,s;\kappa) f(s) ds = \int_{-1}^1 \frac{\kappa}{\pi}\frac{1}{\kappa^2 + (t-s)^2}\,f(s)\,ds
\equiv \int_{-1}^1 Q(t,s;\kappa)\,f(s)\,ds. \]

Let us consider the system \eqref{eqaccoppiatediv1} for $V_1=1, V_2=0$, i.e. \eqref{f1f2}. Substituting the second equation in the first we have
\be 1 = f_1(t) - \int_0^b K(t,s;\kappa) ds \int_0^1 K(s,x;\kappa) f_1(x)\,dx \label{appb1.1}\ee
We can greatly simplify the analysis by adding and subtracting the integral in $s$ in $(b,\infty)$. Using
\be \int_0^\infty K(t,s;\kappa) K(s,x;\kappa) ds = K(t,s;2\kappa)\label{appb1.2}\ee
 we have from \eqref{appb1.1}
\be 1 = f_1(t) - \int_0^1 K(t,x;2\kappa) f_1(x) dx + \int_b^\infty K(t,s;\kappa)ds \int_0^1 dx K(s,x;\kappa) f_1(x) \,.
 \label{appb1.3}\ee
One recognizes in the first three terms the usual Love equation  with scale $2\kappa$. The last term vanishes for $b\to\infty$.
So this calculation confirm the results recollected at the beginning of the section putting them on a more formal basis.

Defining
\be f_1(t) = f_L(t;2\kappa) + \delta f_1(t) \label{appb1.4}\ee
and substituting in \eqref{appb1.3} we obtain the exact equation for $\delta f_1(t)$
\begin{multline}
0 = {\delta f}_1(t) - \int_0^1 dx K(t,x;2\kappa) {\delta f}_1(x) + \int_b^\infty ds K(t,s;\kappa) \int_0^1 dx K(s,x) {\delta f}_1(x) \\
+ \int_b^\infty ds K(t,s;\kappa) \int_0^1 dx K(s,x) f_L(x;2\kappa) \,.\label{numeq1.bis5}
\end{multline}

The solution of \eqref{numeq1.bis5} gives the correction to the solution 
of the disk-plane problem when the plane reduces to a disk larger than the first one.
It is interesting to calculate that correction in the limit $\kappa\to 0$ to find the correct expression of Kirchhoff's formula for the case of different disks.

In fact this equation can be easily solved at the leading order in $\kappa$. In effect due to integration limits $ s\geq b > 1$ the kernels $K$ in the last two terms are non singular in $\kappa$ and the leading term in the solution is obtained replacing $f_L$ by its leading form
\be f_L(t;2\kappa) = \frac{1}{2\kappa} \sqrt{1-t^2}\label{numeq1.bis6}\ee
and neglecting the second-last term in \eqref{numeq1.bis5}. First we note that
\be I(x;\kappa) = 
\int_0^1 K(s,x;\kappa) \sqrt{1-x^2} dx = {\rm Re}[\sqrt{1- (s+ i \kappa)^2} - \kappa]
\mathop{\sim}_{\kappa\to 0} \kappa\left( \dfrac{s}{\sqrt{s^2-1}} - 1\right)\,.
\label{ixk.1}\ee
The approximation is valid for $s\geq b >1$. The equation for $\delta f_1$ is then approximated by
\be \delta f_1(t) = \int_0^1 K(t,x;2\kappa) \delta f_1(x) dx + h(t) \equiv \int_{-1}^1 Q(t,x;2\kappa) \delta f_1(x) dx + h(t)
\label{ixk.2}\ee
with
\[ h(t) = -\frac1{2\kappa} \int_b^\infty ds\, K(t,s) I(s;\kappa)\,. \]
Equation \eqref{ixk.2} has the general form considered in ref.\cite{kac} 
\[ F(t) = \int_{-1}^1 Q(t,s;\kappa) F(s) ds + h(t) \]
where it is shown that the solution in the limit $\kappa\to 0$ has the form
\be F(t) = \frac{1}{\kappa} \int_{-1}^1 ds \,{\cal L}(t,s) h(s) \ee
where
\[ {\cal L}(t,s) = \frac{1}{2\pi} \log\dfrac{1-s t + \sqrt{1-s^2}\sqrt{1-t^2}}{1-s t - \sqrt{1-s^2}\sqrt{1-t^2}} \] 
In our case, with scale $2\kappa$, we have
\be \delta f_1(t) = - \frac1{4\kappa^2} \int_{-1}^1 ds {\cal L}(t,s) \int_b^\infty dx K(s,x;\kappa) I(x;\kappa)\label{appB2.1}\ee
As $s>1$ we can neglect the $\kappa^2$ term in the denominator of $K(t,s;\kappa)$ and perform the resulting integral
\be \int_{-1}^1 ds\,{\cal L}(t,s) K(s,x;\kappa)
\simeq \frac{\kappa}{\pi}\sqrt{\dfrac{1-t^2}{x^2-1}}\left(\frac1{x-t}+ \frac1{x+t}\right)\label{intkp1}\ee
Substituting in \eqref{appB2.1} and performing the last integral in $x$ we have
\begin{multline}
\delta f_1(t) = 
\frac1{2\pi}\dfrac{1}{\sqrt{1-t^2}}\left(t\,{\rm arctanh}\frac{t}{b} - {\rm arctanh}\frac1b\right)\\
+\frac1{4\pi}
\left[\arctan\dfrac{1+ b t}{\sqrt{(b^2-1)(1-t^2)}} + \arctan\dfrac{1- b t}{\sqrt{(b^2-1)(1-t^2)}} \right]
\label{appB2.2}
\end{multline}
Integrating we have for $C_{11}$
\begin{align}
C_{11} &= \frac{2 a}{\pi}\int_0^1 f_1(t) dt = \frac{2 a}{\pi}\int_0^1 (f_L(t;2\kappa) + \delta f_1(t)) dt 
=C_{11}^{(\infty)} + \frac{2 a}{\pi}\int_0^1  \delta f_1(t) dt
\nonumber\\
&
= C_{11}^{(\infty)} + \frac{a}{\pi} \left[  b - \sqrt{b^2-1} -\frac12 {\rm arctanh}\frac1b\right]
= C_{11}^{(\infty)} + \frac{a}{\pi} \left[  b - \sqrt{b^2-1} -\frac14\log\frac{b+1}{b-1}\right]
\label{valoredC11funz}
\end{align}

We note that $\delta f_1(t) = {\cal O}(1)$, in $\kappa$, then neglecting the second-last term in \eqref{numeq1.bis5} shows to be consistent.
Let us observe that for simply computing $\delta C_{11}$ we  could have used
\[ \int_{-1}^1 {\cal L}(t,s) dt = \sqrt{1-s^2} \]
to write from \eqref{appB2.1} (with the usual extension by parity for the range of functions and some manipulations): 
\be \delta C_{11} = \frac{2a}{\pi}\int_{0}^1 \delta f_1(t)dt =
-\frac{2 a}{\pi} \frac1{4\kappa^2} \int_b^\infty ds I^2(s;\kappa)\,.
\label{alterndc11}\ee
Using \eqref{ixk.1} and performing the integral we obtain again \eqref{valoredC11funz}.

To compute $f_2(t)$ it is simpler to consider separately the two intervals $0<t<1$ and $b>t>1$. In the first region we can apply the general
property (see ref.\cite{kac}) 
\be \lim_{\kappa\to 0} \frac1{\kappa}\left\{ \int_{-1}^1 \frac{\kappa}{\pi}\dfrac{g(y)}{\kappa^2 + (x-y)^2} dy\;- g(x)\right\}
= \frac1\pi\dashint \dfrac{g'(y)}{y-x}\,dy \label{lemma}\ee
to eq.\eqref{eqaccoppiatediv1} and write, for small $\kappa$
\[ f_2(t) = - \int_0^1 ds K(t,s;\kappa) f_1(s) = - \int_{-1}^1 ds Q(t,s;\kappa) f_1(s)
\sim - f_1(t) -\frac{\kappa}{\pi}\dashint \dfrac{f_1'(s)}{s-t} ds 
\]
The integral in \eqref{lemma} is a principal value integral.
In the last term we can use the asymptotic form in $\kappa$, $f_1\sim \sqrt{1-s^2}/2\kappa$ and we obtain
\be f_2(t) \sim -f_1(t) + \frac12\,.\ee
For $t>1$ we have, once again, the depression in the kernel and obtain, using \eqref{ixk.1}:
\[ f_2(t) \sim -\frac1{2\kappa}\int_0^1ds\, K(t,s;\kappa) \sqrt{1-s^2} = - \frac{1}{2\kappa} I(t;\kappa)
\simeq -\frac12\left(\dfrac{t}{\sqrt{t^2-1}} - 1\right)\,.\]
Collecting the two contributions, in the bulk region the asymptotic form of $f_2$ is given by
\be f_2(t)\sim - f_1(t)\theta(1-t) + \frac12 - \frac12\dfrac{t}{\sqrt{t^2-1}}\,\theta(t-1) \,.\label{f2asnt}\ee
As usual we remember that we are computing functions in the ``bulk region'', excluding an interval of order $\kappa$ near $t=1$, so
the divergence in \eqref{f2asnt} is only due to this approximation.
The integration in $t$ gives
\[ C_{12} = \frac{2 a}{\pi} \int_0^b f_2(t) dt =
- C_{11} + \frac{a}{\pi} \left[ b - \int_1^b \frac{t}{\sqrt{t^2-1}} dt\right] = - C_{11} + \frac{a}{\pi}
\left(b - \sqrt{b^2-1}\right)\,,\]
equivalent to \eqref{cvsc12}.

A similar analysis can be done for the second independent system, with $V_1 = 0, V_2 = 1$, i.e. for smaller disk at null potential:
\be
0 = g_1(t) + \int_0^b K(t,s;\kappa) g_2(s)ds\,;\quad 1 = g_2(t) + \int_0^1 K(t,s;\kappa) g_1(s)ds\,.
\label{eqg1g2.1}
\ee
These solutions allow us to compute $C_{21}$ and $C_{22}$ and to verify explicitly that $C_{21}=C_{12}$.
Substituting the second equation in the first we have
\[
0 = g_1(t) + \int_0^b K(t,s;\kappa)ds - \int_0^b ds K(t,s;\kappa)\int_0^1 K(s,x;\kappa) g_1(x) dx\,.
\]
The first integral is
\be \int_0^b K(t,s) ds = \frac1\pi \left(\arctan\frac{b-t}{\kappa} + \arctan\frac{b+t}{\kappa}\right)\equiv G_0(t;\kappa,b) 
\,.\label{defG0}\ee
Manipulating the integration limits as above we obtain
\begin{multline*}
0 = g_1(t) +1 + (G_0(t;\kappa,b) - 1) - \int_0^\infty ds K(t,s;\kappa)\int_0^1 K(s,x;\kappa) g_1(x) dx \\ +
\int_b^\infty ds K(t,s;\kappa)\int_0^1 K(s,x;\kappa) g_1(x) dx\\
=1 +  g_1(t) - \int_0^1 dx K(t,s;2\kappa) g_1(x) + (G_0(t;\kappa,b) - 1) + \int_b^\infty ds K(t,s;\kappa)\int_0^1 K(s,x;\kappa) g_1(x) 
dx\,.
\end{multline*}
The first three terms alone are the Love equation with scale $2\kappa$, with opposite sign, the other terms are depressed in $\kappa$ as $\kappa\to 0$. Defining
\be g_1(t) = - f_L(t;2\kappa) + \delta g_1(t) \label{defdg1.1}\ee
we have the exact equation
\begin{multline}
0 = \delta g_1(t) - \int_0^1 dx K(t,s;2\kappa) \delta g_1(x)
 + \bigl( G_0(t;\kappa,b)-1\bigr)\\
  - \int_b^\infty ds K(t,s;\kappa)\int_0^1 dx\,K(s,x;\kappa) ( f_L(x;2\kappa) - \delta g_1(x))\,.
\end{multline}
With the same procedure used above in the last term we can approximate $f_L - \delta g_1$ by their leading term in $\kappa$. Then, we can write the solution at once:
\be
\delta g_1(t) = \frac1{2\kappa}\int_{-1}^1
{\cal L}(t,s)\left[ -\bigl(G_0(s;\kappa,b)-1\bigr) + \frac{1}{2\kappa}\int_b^\infty dx K(s,x) I(x,\kappa)\right]\,ds\,.
\ee
The last term is identical the one in $\delta f_1$, with opposite sign, while the first term is, using the Taylor expansion in $\kappa$ for 
$G_0$:
\be
 - \frac1{2\kappa}\int_{-1}^1
{\cal L}(t,s)\bigl(G_0(s;\kappa,b)-1\bigr)ds = \int_{-1}^1{\cal L}(t,s)\dfrac{b}{\pi}\frac1{b^2-s^2}\,ds
= \frac{1}{\pi} \arctan\sqrt{\dfrac{1-t^2}{b^2-1}}\,.
\ee
Then
\be \delta g_1(t) = - \delta f_1(t) + \frac{1}{\pi} \arctan\sqrt{\dfrac{1-t^2}{b^2-1}}\,.
\label{dg1val.3}\ee
Integrating in $t$ we can compute $C_{21}$
\begin{multline}
C_{21} = \frac{2a}\pi \int_0^1 g_1(t) dt = - \int_0^1 (f_L(t;2\kappa) + \delta f_1(t))dt + \frac{2a}{\pi}\int_0^1 
\frac{1}{\pi} \arctan\sqrt{\dfrac{1-t^2}{b^2-1}}\,dt\\
= - C_{11} + \frac{a}{\pi}\left[b - \sqrt{b^2-1}\right] = C_{12}\,.
\end{multline}
For $g_2$ we can use the same procedure adopted for $f_2$ and we obtain
\be
g_2(t) = \frac12 - g_1(t)\theta(1-t) + \frac12\dfrac{t}{\sqrt{t^2-1}}\theta(t-1)\,. \label{valoreg2}\ee
By integration we obtain \eqref{cvsc12}:
\begin{multline}
C_{22} = \frac{2a}{\pi}\int_0^b g_2(t)dt =
 - C_{12} + \frac{ab}{\pi} + \frac{2a}{\pi}\frac12\int_1^b \dfrac{t}{\sqrt{t^2-1}} dt
 = -C_{12} + \frac{a}{\pi}\left[b + \sqrt{b^2-1}\right]\,.
\end{multline}
\subsection{Interchange of the limits $\kappa\to 0$ and $b\to 1$}
Finally let us make some comments on the crossover region, i.e. what happens if we interchange the limits $b\to 1$ and $\kappa\to 0$.
A look to eq.\eqref{valoredC11funz} brings the suspicion that for $b \sim 1 + \alpha k$, where $\alpha$ is  a numerical factor, 
the expected coefficient $a/4\pi \log(1/k)$ of the logarithmic singularity in the case of equal disks (see eq.\eqref{cijdischiuguali}),
can be reproduced. 
In effect:
\[ C_{11}^{(\infty)} -\frac1{4\pi}\log\frac1\kappa + {\cal O}(1) = \frac{a}{4\kappa} + \frac{a}{\pi}(\frac12 - \frac14) \log\frac1\kappa +
{\cal O}(1) = C_{11}^{eq.disks} + {\cal O}(1)\,.
\]
The same holds for other coefficients $C_{ij}$. This expectation can be explicitly realized by a rough approximation to the limit $b\to 1$.
Consider again $C_{11}$.
Extracting the real part in \eqref{ixk.1} we have, for $x>1$ and $\kappa>0$:
\begin{multline*} I(x;\kappa) = - \kappa + \frac1{\sqrt2}\sqrt{\sqrt{(x^2 - \kappa^2 - 1)^2 + 4\kappa^2 x^2} - (x^2 - \kappa^2 - 1)}\\
= -\kappa + \frac1{\sqrt2}\dfrac{2\kappa x}{
\sqrt{\sqrt{(x^2 - \kappa^2 - 1)^2 + 4\kappa^2 x^2} + (x^2 - \kappa^2 - 1)} }\,.
\end{multline*}
For $\kappa = 0$ in the denominator we get approximation \eqref{ixk.1}. The argument of the square root in $I(x;k)$ is
$2(x^2-1)$ for $\kappa=0$ and $2 \kappa$ for $x=1$, then a rough approximation would be
\[ I(x;\kappa) \simeq \kappa\left( \frac{x}{\sqrt{x^2-1 +\kappa}} - 1\right)\,.\]
Inserting this approximation in  \eqref{alterndc11} we have,, for small $\kappa$:
\begin{multline*} 
\int_{-1}^1 \delta f_1(t) dt \simeq -\frac12 \int_b^\infty
\left( \frac{x^2}{x^2-1 +\kappa} + 1 - \dfrac{2x}{\sqrt{x^2-1+\kappa}}\right) dx\\
\simeq -\frac12 \int_b^\infty dx \left( \frac{1}{x^2-1 +\kappa} + 2 - \dfrac{2x}{\sqrt{x^2-1+\kappa}}\right) 
=  \left( b - \sqrt{b^2-1} + \frac1{4\sqrt{1-k}}\log\dfrac{b-\sqrt{1-\kappa}}{b+\sqrt{1-\kappa}}\right)
\end{multline*}
and for small $\kappa$
\be \delta C_{11} \sim \frac{a}{\pi} \left(b - \sqrt{b^2-1} + \frac1{4}\log\dfrac{b-\sqrt{1-\kappa}}{b+\sqrt{1-\kappa}}\right)\,.
\label{dc11approx}\ee
For $\kappa \to 0$ we find the result  \eqref{valoredC11funz}, but if instead we perform the limit $b\to 1$ we have
\[ \delta C_{11} = \frac{1}{4\pi} \left(4 - 2\log(2) +\log(k) \right)\,. \]
The values of the constants depend on our poor approximation for $I(x;\kappa)$, but the reader can verify that adding the $\log(k)$
term to $C_{11}^{(\infty)}$ one rediscovers the correct behavior \eqref{cijdischiuguali} for equal disks.

\section{Conclusions}
In this paper we reduce the computation of electrostatic potential for  a couple of different circular coaxial disks to a couple of integral equations, generalizing the previous result of Love\cite{Love} for equal disks. The full capacitance matrix is evaluated in the short distance limit, obtaining a generalization of classical Kirchhoff\cite{Kirchh} result. We think that this result could also have some practical applications when a precise knowledge of fringe effects is needed.


\begin{thebibliography}{99}

\bibitem{maxw1} Maxwell, J. C. 1892 \textit{A treatise on electricity and magnetism}, Vol.1, Clarendon Press.
\bibitem{Kirchh} Kirchhoff, G. 1877 \textit{Monatsber. der Akad. der Wiss. zu Berlin} pp. 101-121.
\bibitem{thom}  Thomson, J. J. 1909 \textit{Elements of the Mathematical Theory of Electricity and Magnetism}, Cambridge University Press.
\bibitem{nishi}Nishiyama, H. and Nakamura, M. 1990 Capacitance of Disk Capacitors.
\textit{IEEE Trans. Compon., Hybrids, Manuf. Technol.} \textbf{13}, 417-423.
\bibitem{Muro}
Murovec, T. and Brosseau C. 2014 Numerical simulation of the sign switching of the electrostatic force between charged conducting particles from repulsive to attractive. \textit{J. Appl. Phys.} \textbf{116}, 214902, 1-10.
\bibitem{Qin}
Qin, J., Krapf, N.W., Witten, T.A. 2016 Singular electrostatic energy of nanoparticle clusters. \textit{Phys. Rev. E} \textbf{93}, 022603.
\bibitem{Love}
Love, R. 1949 The electrostatic field of two equal circular co-axial conducting disks. \textit{Quart. J. Mech. Appl. Math.} \textbf{2}, 428-451.
\bibitem{lek}
Lekner, J. 2011 Capacitance coefficients of two spheres. \textit{Journal of Electrostatics} \textbf{69}, 11-14.
 \bibitem{nick}Nicholson, J. 1924 Oblate spheroidal harmonics and their applications. \textit{Philos. T. R. Soc Lon. A} \textbf{224}, 49-93.
\bibitem{sne}Sneddon, I. N. 1966 \textit{Mixed boundary value problems in potential theory}, North-Holland.
\bibitem{Carlson} Carlson, G. and Illman, B. 1994 The circular disk parallel plate capacitor. \textit{Am. J. Phys.} \textbf{62}, 1099-1105.
\bibitem{Hutson}Hutson, V. 1963 The circular plate condenser at small separations.
\textit{Math. Proc. Cambridge Philos. Soc}, {\bf 59}, 211-224.
\bibitem{LL} Landau, L. and Lifshitz, E. 1984 \textit{Electrodynamics of Continuous Media}, Pergamon.
\bibitem{maxw2}Maxwell, J. C. 1879 \textit{The Electrical Researches of Honourable Henry Cavendish, FRS: Written Between 1771 and 1781, edited from the Original Manuscripts}, Note 11, pag. 387-393, Cambridge University Press.
\bibitem{mac} Maccarrone, F. and Paffuti, G. 2016 Capacitance and potential coefficients at large distances. \textit{Journal of Electrostatics}, in press.
\bibitem{kac} Kac, M. and Pollard, H. 1950 The distribution of the maximum of partial sums of independent random variables. \textit{Canadian J. Math.} \textbf{2}, 375-384.
\end{thebibliography}
\end{document}